\documentstyle[aps,epsfig]{revtex}
\begin{document}
\draft

\begin{title}
{\Large\bf Effect of classically forbidden momenta
in one dimensional quantum scattering}
\end{title}
\author{\large A. D. Baute$^{1,2}$, I. L. Egusquiza$^1$ and J. G. Muga$^2$}
\address{
${}^1$ Fisika Teorikoaren Saila,
Euskal Herriko Unibertsitatea,
644 Posta Kutxa, 48080 Bilbao, Spain\\ 
${}^2$ Departamento de Qu\'\i mica-F\'\i sica, 
Universidad del Pa\'\i s Vasco, Apartado 644, 48080 Bilbao, Spain
}
\date{July 20, 2000}
\maketitle

\begin{abstract}
The transmitted wave that results from a collision of a
wave packet which is initially to the left of a potential barrier 
depends in general on the amplitudes of negative momenta of the
initial state.
The exact form of this dependence is shown and the importance of this 
classically forbidden effect is illustrated with numerical 
examples. Special care is taken to account properly for bound states.
\end{abstract}
\pacs{PACS: 03.65.-w
\hfill  EHU-FT/0008}

\section{Introduction}
\begin{figure}[tbp]
    \centerline{\epsfbox{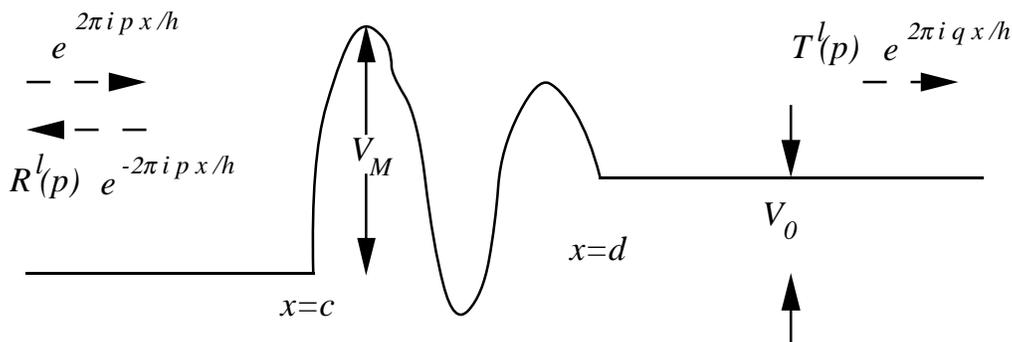}}
    
    \caption{Scattering process in a ``cut-off'' potential: the support 
	of the potential is compact; an incident wave with momentum $p$ 
	produces a transmitted and a reflected wave.}
	\label{fig1}
\end{figure}

Suppose that a classical ensemble of independent particles in one 
dimension is initially 
confined (at $t=0$) in the spatial interval $a<x<b\le 0$, and  
allowed to move freely after $t=0$.
Only particles with positive momenta may arrive at positive positions for 
$t>0$. In contrast, the quantum wave function
involves negative-momentum contributions as well, 
\begin{equation}
\psi(x,t)= h^{-1/2} \int_{-\infty}^\infty dp\,e^{i p x/\hbar} 
\tilde\psi(p) e^{-iE_pt/\hbar},
\end{equation}
where $E_p=p^2/2m$, and 
\begin{equation}
\tilde\psi(p)=h^{-1/2}\int_{-\infty}^\infty dx\,e^{-i p x/\hbar} \psi(x,0)
\end{equation}
is the momentum representation of the initial state.
The effect of negative momenta for $x,t>0$ is however a transient one; the 
total final probability to find the particle at $x>0$ is given only by positive 
momentum components,
\begin{equation}
P_T(\infty)
\equiv\lim_{t\to\infty}\int_0^\infty dx\,|\psi(x,t)|^2=\int_0^\infty dp\,
|\tilde\psi(p)|^2,
\end{equation}
see e.g. \cite{JW} or \cite{Allcock}, since there are no bound 
states. 
This negative-momentum effect is also
present in collisions, where it is combined with other 
classically forbidden effects. 
Consider the family of cut-off potentials of the form 
\begin{equation}
\label{pot}
V(x)=\cases{0 &if $\,\, x<c$\,,\cr
U(x), &if $\,\, c\le x\le d$\,,\cr
V_0,&if $\,\, x>d$,\cr}
\end{equation} 
where $V_0\ge 0$ and $V(x)$ are real, as depicted in Fig. \ref{fig1}. 
%
Let us suppose that the maximum 
value of the potential is $V_M$.
For a classical ensemble confined between $a$ and $b$ (such that 
$a< b\leq c\leq d\le 0$)
only particles with {\it initial} momentum above the ``barrier momentum''
$p_M\equiv(2mV_M)^{1/2}$ may pass to the right of the potential region. 
In the quantum case however, there are contributions
from all the components of the initial wave packet: 
(a) $p>p_M\equiv(2mV_M)^{1/2}$
(above-the-barrier transmission); (b) $p_0<p<p_M$
(``asymptotic'' tunneling, these momenta contribute to 
the transmission probability $P_T(\infty)$); (c) $0<p<p_0$ (``transient tunnelling'',
these momenta do not contribute to $P_T(\infty)$); 
(d) $-\infty<p<0$ (transient negative-momentum effect);
(e) $p_j=i\gamma_j$, $\gamma_j>0$ 
(tunnel effect associated with bound states, which do contribute to $P_T(\infty)$).  

While the various tunnel effects have been well discussed in the 
literature, even though not as clearly distinguished as in the above 
classification, the negative momentum effect, actually the only one 
that survives for free motion, has been frequently overlooked, a 
clear exception being \cite{HWZ,WH}.  This paper is complementary to 
the more usual treatments of one dimensional scattering 
\cite{Newton80a,AKM,RPT,MS}, and lies in the wake of \cite{HWZ,WH}, 
who did consider in a concise manner all these effects, with special 
emphasis on the inclusion of bound states.  Our S-matrix treatment is more 
explicit than that of \cite{HWZ,WH}, and shows how the resolution 
of the identity in scattering eigenstates leads to 
compact expressions for the transmitted wave 
packet. We also illustrate the negative momentum effect with some numerical 
examples and work out in detail both the case when the state is 
confined initially to the lower level and when the initial confining 
is to the upper level (for which case the frequently overlooked 
contributions are those of positive momenta, including evanescent 
waves).  The contribution of bound 
states is indicated explicitly, both formally and with some numerical examples.  
\section{``Stationary'' eigenstates of the Hamiltonian}
The total Hamiltonian $H=H_0+V$ may have a discrete set of bound states 
$\{|E_j\rangle\}$, with energies $E_j<0$ and (real) wavefunctions $\phi_{j}(x)$,  
and a continuum of ``stationary  scattering eigenstates'' with $E_p>0$. 
Only the former belong to the Hilbert
space of square integrable functions. The latter however form a
convenient basis normalized according to Dirac's delta. 
For energies above $V_0$, the energy spectrum is doubly degenerate 
as corresponds physically to incidence from one side or the other. 
Below $V_0$ there is only one linearly independent 
solution.  
The resolution of the identity may be written in different ways, in
particular as 
\begin{equation}\label{1}
1=\sum_j|E_j\rangle \langle E_j| +\int_{-\infty}^{-p_0}dp\, 
|p^\pm\rangle\langle p^\pm|
+\int_{p_{0}}^\infty dp\,|p^\pm\rangle\langle p^\pm|
\pm\int_{0}^{\pm p_{0}}dp\,|p^\pm\rangle\langle p^\pm|,
\end{equation}
where the states $|p^\pm\rangle$ have energy $E_p=p^2/(2m)$. 
Continuum and bound states are orthogonal. 
Moreover, $\langle p^\pm| {p'}^\pm\rangle=\delta(p-p')$, and 
$\langle E_j|E_j'\rangle=\delta_{jj'}$. 
The first two integrals in Eq.~(\ref{1}) reflect the double degeneracy 
signalled above, whereas the last integral corresponds to the 
non-degenerate part of the continuous spectrum. The reason for the 
sign choice in front of the last integral is due to the unified 
notation we shall now introduce for the generalized eigenstates 
associated with the continuous part of the spectrum.

The states $|p^+\rangle$ with $p>0$ are characterized by an 
{\it incident} plane wave of momentum $p$ from the left. 
The asymptotic behaviour of their wave-functions $\psi_{p^{+}}(x)$ is 
\begin{equation}\label{pl}
\psi_{p^{+}}(x)=
\frac{1}{h^{1/2}}\times\cases{\exp(ipx/\hbar)
+R^l(p)\exp(-ipx/\hbar),& $\,\, x<c$\cr
T^l(p)\exp(iqx/\hbar),& $\,\, x>d$,\cr}
\end{equation}
where $q$ is the momentum with respect to the upper (right) level
of the potential, $q=(p^2-2mV_0)^{1/2}$. For the evanescent regime,
i.e. when  $p^2<2mV_0$, the plane waves on the right become decaying
exponentials so the positive imaginary square root is taken.    
$R^l(p)$ and $T^l(p)$ are reflection and transmission amplitudes for 
left incidence. They are obtained by solving the stationary Schr\"odinger 
equation subject to the boundary conditions specified in (\ref{pl}).  

The states $|p^+\rangle$ for $p<-p_0$ are defined by having an
{\it incident} plane wave 
from the right, oscillating with spatial frequency $|q|/\hbar$, 
where now the negative square root is taken, 
$q=-(|p^2-2mV_0|)^{1/2}$.  
The asymptotic behaviour of the corresponding wave-functions is 
\begin{equation}\label{pr}
\psi_{p^{+}}(x)=\frac{1}{h^{1/2}}\left(\frac{p}{q}\right)^{1/2}\times
\cases{T^r(-p)\exp(ipx/\hbar),& $\,\, x<c$\cr
\exp(iqx/\hbar)+R^r(-p)\exp(-iqx/\hbar),& $\,\, x>d$,\cr}
\end{equation}
The factor $(p/q)^{1/2}$ is necessary for the proper delta normalization. 
$R^r(-p)$ and $T^r(-p)$ are reflection and transmission amplitudes for 
right incidence. Note that the arguments of transmission or reflection 
amplitudes are always positive for states $|p^+\rangle$, independently of the sign of 
$p$. 

In both cases, (\ref{pl}) and (\ref{pr}), 
$q$ may be defined as the square root of $p^2-2mV_0$
with a branch cut that joins the branch points $p=\pm p_0$ 
going slightly below ${\rm Im}(p)=0$.  In this way the sign of $q$ is the same 
as the sign of $p$ for $p^2>p_{0}^2$.

The states $|p^-\rangle$ are obtained simply by considering $p<0$ 
in expression (\ref{pl}) and $p>p_{0}$ in expression (\ref{pr}). 
In these states there appears
an outgoing plane wave instead of an incident 
plane wave. Thus, the right hand side of Eq. (\ref{pl}), with $p<0$, defines the 
corresponding 
states $|p^-\rangle$, as those with a left-outgoing plane wave 
of absolute momentum $|p|$, and can be read as $\psi_{p^{-}}(x)$, 
for $p<0$. For $p<-p_0$, $q<0$, whereas for 
$-p_0<p<0$, $q$ becomes as before a positive imaginary number.      
States $|p^-\rangle$ with a right-outgoing plane wave are defined by the 
right hand side of Eq. 
(\ref{pr}), with $p>p_0$, and $q>0$. Note that the arguments of the
amplitudes $R^{r,l}(p)$ and $T^{r,l}(p)$ are negative for $|p^-\rangle$ states. 
Since the formal boundary conditions that define
the states $|p^\pm\rangle$ are in fact equal, the negative-argument
amplitudes will be given by the same formal expressions valid for their 
positive-argument counterparts. For the same reason,
we shall refer to $R^{r,l}(p)$ and $T^{r,l}(p)$
as ``reflection'' and ``transmission'' amplitudes independently of
the sign of $p$, even though, on physical grounds, this terminology 
and notation would only be appropriate for 
the $p>0$ case (only in that case does $T^{r,l}$  multiply a
{\it transmitted} wave  
and $R^{r,l}$ a {\it reflected} wave).

Reflection and transmission amplitudes are not independent.
The unitarity of the ${\sf S}$ 
matrix imposes certain relations among them. 
The ${\sf S}$ matrix elements are defined as the 
coefficients multiplying the outgoing plane waves 
when the incident plane wave is normalized to unit flux.
When the two channels of the one dimensional scattering 
are open (this happens for $p>p_0$), they are given by 
\begin{eqnarray}
{\sf S}(p)=\left(
\begin{array}{cc}
\left(\frac{q}{p}\right)^{1/2}T^l(p)&R^l(p)
\\
R^r(p) &\left(\frac{p}{q}\right)^{1/2}T^r(p) \\
\end{array}
\right)
\end{eqnarray}
The unitarity of the ${\sf S}$ matrix, ${\sf S}{\sf S}^\dagger={\bf 1}$, implies
that 
\begin{eqnarray}\label{uni1}
\frac{p}{q}|T^r(p)|^2+|R^r(p)|^2=1\,,
\\
\label{uni2}
|R^l(p)|^2+\frac{q}{p}|T^l(p)|=1\,,
\\
\label{uni3}
\frac{p}{q}T^r(p)R^l(p)^*+R^r(p)T^l(p)^*=0\,.
\end{eqnarray}
For $0<p<p_0$ only one channel is open, and the ${\sf S}$ matrix 
reduces to a number, $R^l(p)$. Unitarity implies in this case 
\begin{equation}
\label{uni4}
R^l(p)R^l(p)^*=1,\;\;\;\;\;\; 0<p<p_0\,.
\end{equation}
All these equations, from (\ref{uni1}) to
(\ref{uni4}), are also valid for negative momenta and relate the
amplitudes associated with $|p^-\rangle$ states.  They can also be derived 
by comparing various Wronskians of the stationary 
scattering states at different regions. 

For the evanescent case, $0<p<p_0$, one important relation follows by 
multiplying $|p^+\rangle$ by $R^l(p)^*$ and using (\ref{uni4}). This gives 
the state $|-p^-\rangle$. Equating the coefficients for $x>d$,     
\begin{equation}\label{evan}
T^l(-p)=T^l(p)R^l(p)^*,\;\;\;\;\;0<p<p_0.
\end{equation}
Similarly, by taking the complex conjugate of the boundary
conditions (\ref{pl}) and (\ref{pr}), it is found that 
$\psi_{-p^-}(x)=\psi_{p^+}(x)^*$, 
and comparing the coefficients that multiply the exponentials, 
\begin{eqnarray}
\label{pm}
T^{r,l}(-p)&=&T^{r,l}(p)^*\,,
\\     
R^{r,l}(-p)&=&R^{r,l}(p)^*\,.
\end{eqnarray}
Another important relation between $T^l(p)$ and $T^r(p)$ 
follows by equating the Wronskians of $\psi_{{p^\pm}}(x)$
and $\psi_{{-p^\pm}}(x)$ at $x<c$ and $x>d$, 
\begin{equation}\label{tlr}
T^r(p)p=T^l(p)q.
\end{equation}

As an illustration of the above, the reflection and transmission amplitudes 
for the simple step potential ($c=d=0$, $U(x)=0$) are given, for all $p$, by 
\begin{eqnarray}
T^l(p)&=&\frac{2p}{q+p}\,;\;\;\;\;\;\;R^l(p)=\frac{p-q}{q+p}\,;
\nonumber\\
T^r(p)&=&\frac{2q}{p+q}\,;\;\;\;\;\;\;R^r(p)=\frac{q-p}{p+q}\,.
\label{step}
\end{eqnarray}
The reader may easily check relations (\ref{uni1}-\ref{tlr}) for 
amplitudes that correspond to the step potential, 
Eq. (\ref{step}), as a test of their validity.
\section{A compact expression for the transmitted wave function} 
In this section we shall find, using 
the relations of the previous one, an expression for
$\psi(x,t)$, with $t>0$ and 
$x>d$, assuming that the initial wave function is 
restricted to $a<x<b\leq c$.     
First we insert the resolution of the identity in terms of 
bound states and scattering states $|p^+\rangle$,
\begin{eqnarray}
\psi(x,t)&=&
\sum_j \phi_j(x) \langle E_j|\psi(0)\rangle e^{-iE_j t/\hbar}
\nonumber\\   
&+&\int_{-\infty}^{-p_0} dp\,\psi_{p^+}(x) \langle p^+|\psi(0)\rangle e^{-iE_pt/\hbar}
\nonumber\\
&+&\int_{0}^{\infty} dp\,\psi_{p^+}(x) 
\langle p^+|\psi(0)\rangle e^{-iE_pt/\hbar},\;\;\;\;\;\;
\forall x {\rm~and~} \forall t.
\label{20}\end{eqnarray}
Because of the initial restriction of the wave function, the matrix element 
$\langle p^+|\psi(0)\rangle$ may be evaluated with the aid of Eqs. (\ref{pl}) and 
(\ref{pr}), 
\begin{eqnarray}\label{21}
\langle p^+|\psi(t=0)\rangle=\cases{
\left(\frac{p}{q}\right)^{1/2}\tilde\psi(p) T^r(-p)^*\,, &$p<-p_0$\,,
\cr
\tilde\psi(p)+
\tilde\psi(-p) R^l(p)^*\,,& $p>0$\,.
\cr}
\end{eqnarray}
Now we use Eqs. (\ref{pl}-\ref{pr}) and (\ref{21}) in (\ref{20}), for 
$x>d$,
\begin{eqnarray}
\psi(x,t)&=&
\sum_j \phi_j(x) \langle E_j|\psi(0)\rangle e^{-iE_jt/\hbar}
\nonumber\\   
&+&h^{-1/2}\int_{-\infty}^{-p_0} dp\, 
\frac{p}{q}\left[e^{iqx/\hbar}+R^r(-p)e^{-iqx/\hbar}\right]
T^r(-p)^*\tilde\psi(p) e^{-iE_pt/\hbar}
\nonumber\\
&+&h^{-1/2}\int_0^\infty dp\,
T^l(p)e^{iqx/\hbar}\left[\tilde\psi(p)+R^l(p)^*\tilde\psi(-p)\right]
e^{-iE_pt/\hbar},\;\;\;\;\;\;x>d, 
\end{eqnarray}
and reorganize the terms as follows,
\begin{eqnarray}
\psi(x,t)&=&
\sum_j \phi_j(x) \langle E_j|\psi(t=0)\rangle e^{-iE_jt/\hbar}
\nonumber\\   
&+&h^{-1/2}\int_{-\infty}^{-p_0} dp\, 
\frac{p}{q}T^r(-p)^* \tilde\psi(p) e^{iqx/\hbar}e^{-iE_pt/\hbar}
\nonumber\\
&+&h^{-1/2}\int_{-\infty}^{-p_0} dp\,
\left[\frac{p}{q}R^r(-p)T^r(-p)^*+T^l(-p)R^l(-p)^*\right]
\tilde\psi(p) e^{-iqx/\hbar}e^{-iE_pt/\hbar}
\nonumber\\
&+&h^{-1/2}\int_{-p_0}^0 dp\,T^l(-p)R^l(-p)^* \tilde\psi(p)
e^{iqx/\hbar}e^{-iE_pt/\hbar}
\nonumber\\
&+&
h^{-1/2}\int_{0}^{\infty} dp\,T^l(p)  \tilde\psi(p) e^{iqx/\hbar}e^{-iE_pt/\hbar},
\;\;\;\;\;\;\;x>d.\label{mogollon}
\end{eqnarray}
Keep in mind that between $-p_0<p<p_0$, $q$ is a positive 
imaginary number, $q=i(|p^2-2mV_0|)^{1/2}$. The second term in the third line
and the fourth line come from a variable change $p\to-p$.

Making use of Eqs. (\ref{uni3}-\ref{tlr}), there results
the simple form
\begin{eqnarray}\nonumber
\psi(x,t)&=&
\sum_j \phi_j(x) \langle E_j|\psi(t=0)\rangle e^{-iE_jt/\hbar}
\\   
&+&h^{-1/2}\int_{-\infty}^{\infty} dp\, 
T^l(p) \tilde\psi(p) e^{iqx/\hbar}e^{-iE_pt/\hbar}\,,\;\;\;\;\;\;x>d.
\label{almost}
\end{eqnarray}
The equation can be put in an even more compact form as shown in the 
next section 
\section{Bound states}
None of the particles of the classical ensemble described in the introduction
may be trapped by a potential well of the potential $U(x)$. In quantum mechanics
though, a bound state wave function extends exponentially beyond the potential 
limits, and may overlap with the initial state $\psi(0)$, even when this state 
is localized outside the potential limits. 
The contribution of these bound states to the wave function at $t>0$ 
is orthogonal to the scattering (continuum) part, and will remain spatially
linked to the potential region at all times.  

A bound state with energy $E_j$ corresponds to a simple pole
of $T^l(p)$, or a zero of $1/T^l(p)$ on the positive imaginary axis,
at $p_j=i\gamma_j$, $\gamma_j>0$. 
In this section we shall see that the bound state terms may be written as
a residue 
\begin{equation}\label{resi}
\phi_j(x) e^{-iE_jt/\hbar}\langle E_j|\psi(0)\rangle=-2\pi i h^{-1/2}\, {\rm 
Res}\,
\left[T^l(p) \tilde\psi(p) e^{i q_{j} x/\hbar} e^{-iE_pt/\hbar}\right]_{p=p_j},\;\;
\;\;\;\;x>d
\end{equation}
where $q_{j}=i\sqrt{\gamma_{j}^2+p_{0}^2}$.
The function $\tilde\psi(p)$ is defined on the complex plane by the 
integral
\begin{equation}\label{psipdef}
\tilde\psi(p)=\frac{1}{h^{1/2}}\int_{a}^{b} dx\,
e^{-ipx/\hbar} \psi(x,0), 
\end{equation}
as has been used all along.

\begin{figure}[t]
    \vspace{7mm}
    \centerline{\epsfbox{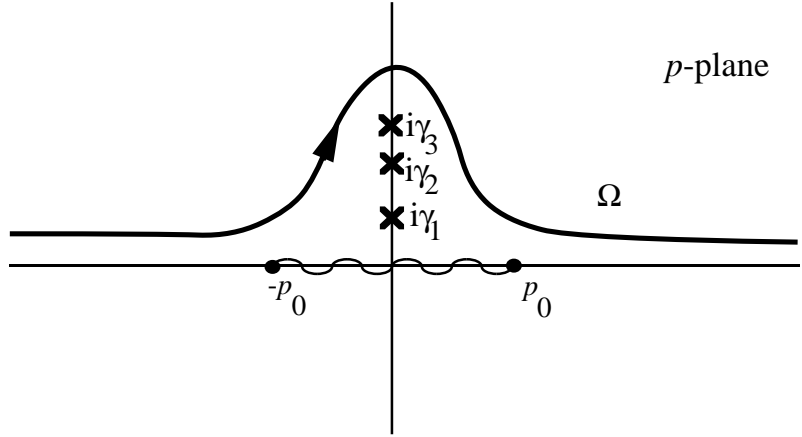}}
    \caption{Path of integration for Eq. (\ref{comp1}) }
	\label{figomega}
\end{figure}

For the formal treatment of bound states it is 
convenient to introduce Jost solutions $f_1(p,x)$ and $f_2(p,x)$
of the Schr\"odinger equation. They are  
defined by the boundary conditions
\begin{eqnarray}
f_1(p,x)&\to&e^{iqx/\hbar}, \;\;\;\;\;\;{\rm as~}x>d\,,\quad {\rm and}
\nonumber\\
f_2(p,x)&\to&e^{-ipx/\hbar},\;\;\;\;\;\;x<c\,.
\end{eqnarray}
We generalize here the treatment of ref.\cite{Newton80a}
for $V_0=0$ to the case $V_0\ge 0$.
It is easy from (\ref{pl}) and (\ref{pr}) to obtain the explicit expressions of 
$f_1$ and $f_2$ at $x<c$ and $x>d$ respectively,  
\begin{equation}\label{jost1}
f_1(p,x)=\cases{
e^{iqx/\hbar}\,, &$\,\,x>d$\cr
e^{ipx/\hbar}\frac{1}{T^l(p)}+e^{-ipx/\hbar}\frac{R^l(p)}{T^l(p)}\,;& $\,\,x<c$\cr}
\end{equation}
\begin{equation}\label{jost2}
f_2(p,x)=\cases{
e^{iqx/\hbar}\frac{R^r(p)}{T^r(p)}+e^{-iqx/\hbar}\frac{1}{T^r(p)}\,, &$\,\,x>d$\cr
e^{-ipx/\hbar}\,.&$\,\, x<c$\cr}
\end{equation}
In particular, 
if for some value $p_j$, $1/T^l(p_j)=0$, then $f_2$ and $f_1$ become real, 
proportional to each other,
\begin{equation}
f_2=Cf_1,
\end{equation}
and decay exponentially for $x<c$ and $x>d$. 
Defining the normalization constant $N$ by 
\begin{equation}
N^2=\int_{-\infty}^\infty dx\, f_1^2\,,
\end{equation}
we may write the wave-function of the bound state $|E_{j}\rangle$ as
\begin{equation}\label{rela}
\phi_{j}(x)={1\over{N}}f_{1}(p_{j},x)={1\over{CN}}f_{2}(p_{j},x)\,.
\end{equation}
The overlap between the bound state $|E_{j}\rangle$ and the initial state, 
which has support only on the interval $[a,b]\leq c$, is as usual
\begin{equation}\label{overlap}
\langle E_{j}|\psi(0)\rangle=\int_{a}^bdx\,\phi_{j}(x)\psi(x,0)\,.
\end{equation}
On substituting the second equality of Eq. (\ref{rela}) in Eq. 
(\ref{overlap}), and taking into account definition (\ref{psipdef}) 
and Eq. (\ref{jost2}), 
we obtain
\begin{equation}\label{pjpsi}
\langle E_{j}|\psi(0)\rangle={1\over{CN}}\int_{a}^bdx\,f_{2}(p_{j},x)\psi(x,0)
={1\over{CN}}\int_{a}^bdx\,
e^{-ip_{j}x/\hbar}\psi(x,0)={{h^{1/2}}\over{CN}}\tilde\psi(p_{j})\,.
\end{equation}
Therefore, for $x>d$, we have
\begin{eqnarray}
\phi_j(x)\langle E_j|\psi(0)\rangle&=&\left({1\over{N}}f_{1}(p_{j},x)\right)\langle 
E_j|\psi(0)\rangle =\left({1\over{N}}e^{i q_{j}x/\hbar}\right)\langle 
E_j|\psi(0)\rangle\nonumber\\
&=&\frac{h^{1/2}}{CN^2}e^{i q_{j}x/\hbar} \tilde\psi(p_{j}) =
-2\pi i h^{-1/2} e^{i q_{j}x/\hbar}\tilde\psi(p_{j}) {\rm Res}\, 
T^l(p)_{p=p_j} \,,
\label{tra}
\end{eqnarray}
so that (\ref{resi}) is obtained. 
In the last line of (\ref{tra}) we have used \cite{Newton80a} 
\begin{equation}
{\rm{Res}}\,T^l(p)_{p=p_j}=
\left[\frac{\partial}{\partial p}(1/T^l(p))\right]^{-1}_{p=p_j}
=\frac{i\hbar}{CN^2},
\end{equation}
a relation that may be obtained by differentiating the Schr\"odinger equation 
for $f_1$ with respect to $p$, integrating
$f_1f_2$ between $-R$ and $R$ as $R\to\infty$, and comparing with the derivative
of the Wronskian of $f_1$ and $f_2$ with respect to $p$ for $p=p_j$, see 
also \cite{GW}.   
Finally, combining (\ref{almost}) and (\ref{resi})
we can write the transmitted wave packet very 
compactly as 
\begin{equation}\label{comp1}
\psi(x,t)= h^{-1/2}
\int_\Omega dp\,T^l(p) 
\tilde\psi(p) e^{i q x/\hbar} e^{-iE_pt/\hbar},\;\;\;\;\;\;\; x>d,
\end{equation}
where the contour $\Omega$ goes from $-\infty$ to $\infty$ passing above the bound
state poles, as shown in Fig.~\ref{figomega}. 
%
\section{Initial state in the upper level}
Imagine now that the initial state has support only to the right, 
where the potential is $V_{0}$; that is, assume that the initial 
support of the state is the interval $[a,b]$, such that $d\leq a<b$. In 
this case, instead of Eq. (\ref{21}), we have
\begin{eqnarray}\label{psiup}
\langle p^+|\psi(t=0)\rangle=\cases{
 T^l(p)^*\tilde\psi(q)\,, &$p>p_{0}$\,,
\cr
 T^l(p)^*\tilde\psi(-q)\,, &$0<p<p_{0}$\,,
\cr
\left(\frac{p}{q}\right)^{1/2}\left(\tilde\psi(q)+
 R^r(-p)^*\tilde\psi(-q)\right)\,,& $p<-p_{0}$\,,
\cr}
\end{eqnarray}
where, as before, $q=\left(p^2-p_{0}^2\right)^{1/2}$, with the branch 
cut going slightly below ${\rm Im}(p)=0$, and $\tilde\psi(q)$ is defined by 
Eq. (\ref{psipdef}).

Expression (\ref{20}), as such, is also valid for $x<c$. Let us then 
substitute Eq. (\ref{psiup}) in Eq. (\ref{20}), and use Eqs. 
(\ref{pl}) and (\ref{pr}). The result analogous to Eq. 
(\ref{mogollon}) reads then, for $x<c$ when the initial state is 
restricted at $x>d$,
\begin{eqnarray}
\psi(x,t)&=&
\sum_j \phi_j(x) \langle E_j|\psi(t=0)\rangle e^{-iE_jt/\hbar}
\nonumber\\   
&+&h^{-1/2}\int_{-\infty}^{-p_0}dp\, T^l(p)^* \tilde\psi(q) 
e^{ipx/\hbar}e^{-iE_pt/\hbar}
\nonumber\\
&+&
h^{-1/2}\int_{p_{0}}^{\infty} dp\,T^l(p)^* \tilde\psi(q) 
e^{ipx/\hbar}e^{-iE_pt/\hbar}\nonumber\\
&+&
h^{-1/2}\int_{0}^{p_{0}} dp\,T^l(p)^* \tilde\psi(-q) 
e^{ipx/\hbar}e^{-iE_pt/\hbar}\nonumber\\
&+&h^{-1/2}\int_{p_0}^{\infty} dp\,
\left[\frac{p}{q}R^r(p)^*T^r(p)+T^l(p)^*R^l(p)\right]
\tilde\psi(q) e^{-ipx/\hbar}e^{-iE_pt/\hbar}
\nonumber\\
&+&h^{-1/2}\int_{0}^{p_{0}} dp\,T^l(p)^*R^l(p) \tilde\psi(-q)
e^{-ipx/\hbar}e^{-iE_pt/\hbar},
\;\;\;\;\;\;\;x<c.\label{mogollon2}
\end{eqnarray}
Making use of the unitarity relations, this expression can be written 
as
\begin{equation}
\psi(x,t) = \sum_j \phi_j(x) \langle E_j|\psi(t=0)\rangle e^{-iE_jt/\hbar}
+ h^{-1/2} \int_{\Gamma}dp\, 
T^l(-p) \tilde\psi(q) 
e^{ipx/\hbar}e^{-iE_pt/\hbar}\,,
\end{equation}
where the path of integration $\Gamma$ goes from $-\infty$ to 
$+\infty$ immediately {\it below} the branch cut. Notice that for 
$p=-i\gamma$, with real positive $\gamma$, $q$ becomes 
$-i\sqrt{\gamma^2+p_{0}^2}$, which accounts for the minus sign in front 
of $q$ in the 
third and fifth integrals of Eq. (\ref{mogollon2}).

In order to include the bound states in a more compact expression,
let us substitute the first equality of Eq. (\ref{rela}) in Eq. 
(\ref{overlap}). Thus, using definition (\ref{psipdef}) 
and Eq. (\ref{jost1}), 
we obtain
\begin{equation}\label{pjpsi2}
\langle E_{j}|\psi(0)\rangle={1\over{N}}\int_{a}^bdx\,f_{1}(p_{j},x)\psi(x,0)
={1\over{N}}\int_{a}^bdx\,
e^{iq_{j}x/\hbar}\psi(x,0)={{h^{1/2}}\over{N}}\tilde\psi(-q_{j})\,,
\end{equation}
where, as before, $q_{j}=i\sqrt{\gamma_{j}^2+p_{0}^2}$.
Therefore, for $x<c$, we have in this situation, using Eqs. 
(\ref{jost2}) and (\ref{rela}),
\begin{eqnarray}
\phi_j(x)\langle E_j|\psi(0)\rangle&=&\left({1\over{CN}}f_{2}(p_{j},x)\right)\langle 
E_j|\psi(0)\rangle =\left({1\over{CN}}e^{-i p_{j}x/\hbar}\right)\langle 
E_j|\psi(0)\rangle\nonumber\\
&=&\frac{h^{1/2}}{CN^2}e^{-i p_{j}x/\hbar}\tilde\psi(-q_{j}) \,. \label{tra2}
\end{eqnarray}
The pole of $T^l(p)$ at $p_{j}=i\gamma_{j}$ becomes a pole of 
$T^l(-p)$ at $-p_{j}$. Thus,
\begin{equation}
T^l(-p)=\frac{-i\hbar}{CN^2}\frac{1}{p+p_{j}} + \cdots\,,
\end{equation}
and it follows that 
\begin{equation}
\phi_j(x) \langle E_j|\psi(t=0)\rangle e^{-iE_jt/\hbar} = 2\pi i h^{-1/2}
{\rm~Res}\left[T^l(-p)e^{i p 
x/\hbar}\tilde\psi(q) 
e^{-iE_{p}t/\hbar}\right]_{p=-p_j}\,,
\end{equation}
from which we can conclude that when the initial state is restricted 
to the interval $[a,b]$, with $d<a<b$,  
\begin{equation}
\psi(x,t) =  h^{-1/2} \int_{\Omega'}dp\, 
T^l(-p) \tilde\psi(q) 
e^{ipx/\hbar}e^{-iE_pt/\hbar}\,,\label{comp2}
\end{equation}
for $x<c$, where $\Omega'$ is a path of integration that goes from $-\infty$ 
to $+\infty$ {\it below} the branch cut and the poles $-p_{j}$.

\section{Discussion}
The main results of this paper are Eqns. (\ref{comp1}) and (\ref{comp2}),
that provide simple, compact, and exact expressions for the transmitted 
wave packet in terms of the initial momentum distribution and bound state 
contributions. Whereas these expressions have been obtained for cut-off potentials,
they may be easily generalized for potentials that decay fast enough, so that 
the initial wave packet does not overlap significantly with the potential 
region, and for $x$ values where the asymptotic expressions for the scattering 
states are accurate.

These two equations make clear the need to include different contributions 
for the transmitted wave-packet. Even though tunnelling terms are of course 
included in theoretical analysis of wave packet collisions, 
the contribution of negative momenta is frequently overlooked in integral expressions 
of the wave function. Similarly, the contribution of the evanescent
momentum region when the wave packet
is initially on the potential upper level has been also disregarded.
We have applied these equations in \cite{BEM} 
to establish the relation between source boundary conditions, in which the wave
function is specified at a point and for all times,  with standard
initial-value-problem boundary conditions where the wave function is specified
in all space at $t=0$.
One further intended application is the study of arrival-time measurement 
models where a clock dial is coupled to the particle's 
motion in such a way that the particle's crossing stops the dial's
motion \cite{AOPRU,ML}. 
These models are described by step potentials of the form considered in section V.    
  
Let us now illustrate the formal results obtained above with some 
simple explicit computations. In all of them,
the initial wave packet is taken as the ground state of an infinite well; it 
is located between $a$
and $b$,
\begin{equation}\label{psiiv}
\psi(x,0)=\left(\frac{2}{b-a}\right)^{1/2}\sin[(x-a)k_w]{\cal H}(a,b),
\end{equation}
where $k_w=\pi/(b-a)$, and
\begin{equation}
{\cal H}(a,b)=\cases{
1&$\,\,\, a<x<b$,\cr
0&$\,\,\, {\rm otherwise}$.\cr}
\end{equation}
See \cite{BEM} for details of the analytical time dependence 
of $\psi(x,t)$. 
Here we shall evaluate the norm for $x>0$,
\begin{equation}
P_T(t)=\int_0^\infty dx\,|\psi(x,t)|^2\,,
\end{equation}
as well as the 
contributions from positive and negative momenta, the interference 
terms, and, whenever required, the contribution of bound states and 
evanescent waves.

%
%
\begin{figure}[t]
    \vspace{7mm}
    \centerline{\epsfbox{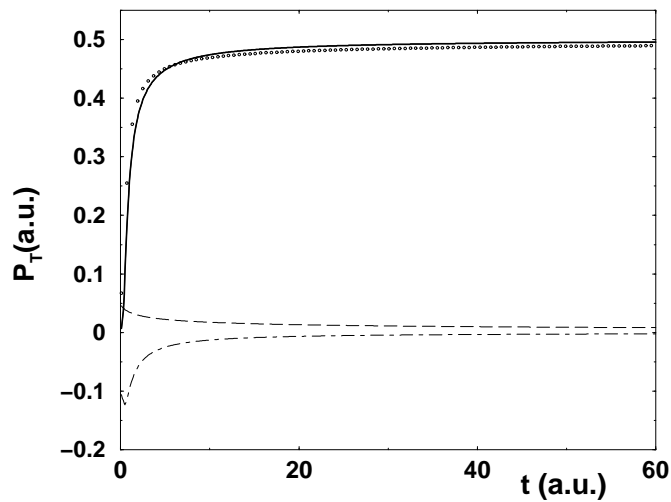}}
	\caption{The contributions of positive and negative momentum  to the probability 
	of finding the particle in the half line $x>0$ for time $t$ in the 
	free case are 
	shown with a dotted and a dashed line, respectively. The interference 
	term is depicted with a dash-dotted line, and 
	the total probability of finding the particle in the half line $x>0$ for 
	time $t$ is portrayed as a thick continuous line. The initial state 
	is given by expression (\ref{psiiv}), with $a=-2.01$, $b=-0.01$. The 
	mass $m$ equals $1$. Atomic units are used throughout.}
	\label{fig2}
\end{figure}
For instance, in the case of free motion, by defining the projectors 
\begin{eqnarray}
P&=&\int_0^\infty dp\,|p\rangle\langle p|\quad{\rm and}
\\
Q&=&1-P\,,
\end{eqnarray}
$P_T$ is decomposed into three terms, $P_T=P_{T,+}+P_{T,-}+P_{T,int}$:  
\begin{eqnarray}
P_{T,+}(t)&=&\int_0^\infty dx\,|\langle x|P\psi(t)\rangle|^2\,,
\noindent
\\
P_{T,-}(t)&=&\int_0^\infty dx\,|\langle x|Q\psi(t)\rangle|^2\,,
\noindent
\\
P_{T,int}(t)&=&\int_0^\infty dx\, 2 {\rm Re}[\langle x|P\psi(t)\rangle 
\langle Q\psi(t)|x\rangle]  \,.
\end{eqnarray}
These quantities are shown as functions of $t$ for free motion in 
Fig.~\ref{fig2}. The interference term is important only for early 
times, and the limit $P_{T}(\infty)=P_{T,+}(\infty)$ is clearly 
checked visually, although the decay of $P_{T,-}$ is rather slow.

\begin{figure}[t]
    \vspace{7mm}
    \centerline{\epsfbox{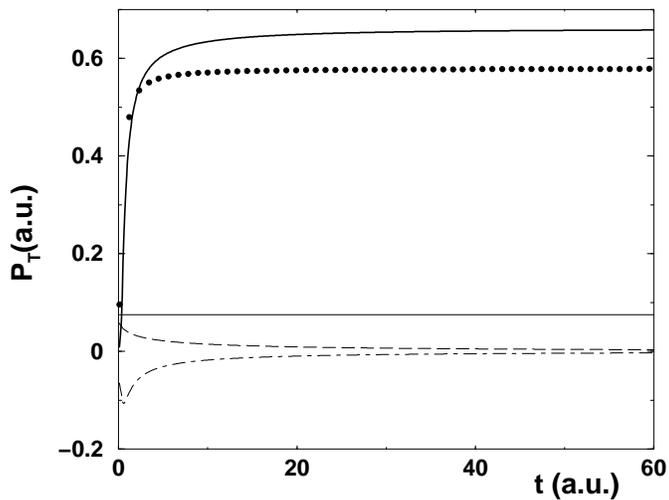}}
	\caption{The initial state is as in Eq.  (\ref{psiiv}), but 
	multiplied by $\exp(ipx)$, where $p=0.5$ (all quantities in atomic 
	units).  Again we have $a=-2.01$ and $b=-0.01$.  The motion of the 
	particle is free everywhere, except for an attractive delta 
	potential, $U(x)=-U_{0}\delta(x)$, with $U_{0}=1/8$.  The 
	contribution of the bound state to the total probability of 
	finding the particle in the half line $x>0$ is shown with a thin 
	continuous line; the contributions of positive and negative momenta 
	with a dotted and a dashed line, respectively.  The interference 
	term between positive and negative momenta is depicted with a 
	dash-dotted line, and the total probability as a thick continuous 
	line.  }
    \label{fig3}
\end{figure}

\begin{figure}[t]
    \vspace{7mm}
    \centerline{\epsfbox{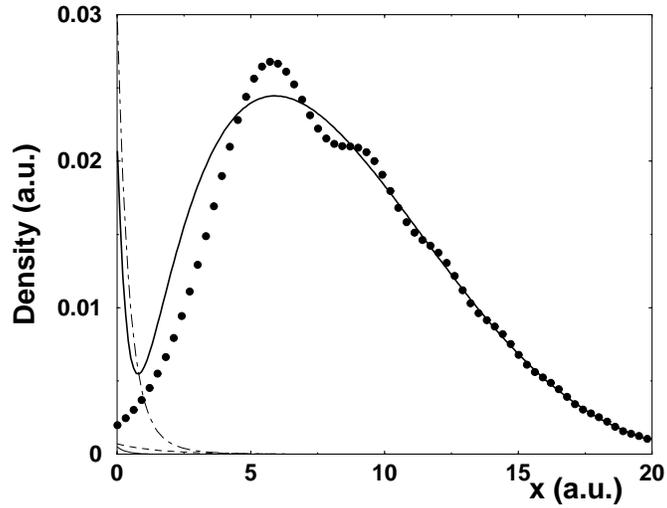}}
    \caption{The particle of mass $m=1$ moves in the potential 
    $V(x)=-U_{0}\delta(x)+V_{0}\theta(x)$, where $U_{0}=1/8$ and 
    $V_{0}=1$ (in atomic units). The initial state is as in Eq.  
    (\ref{psiiv}), with $a=-2.01$ and $b=-0.01$, but 
	multiplied by $\exp(ipx)$, where $p=0.5$. Time $t=5$ has elapsed since 
    the initial state was released. The moduli squares of different 
    contributions to the amplitude are 
    depicted: continuous thick line, total amplitude modulus squared; dotted 
    line (resp. dashed line), square modulus of the contribution to the amplitude 
    of positive  (resp. negative) momenta; dot-dashed line, evanescent 
    waves; continuous thin line, bound state contribution.}
    \label{fig4}
\end{figure}

\begin{figure}[t]
    \vspace{7mm}
    \centerline{\epsfbox{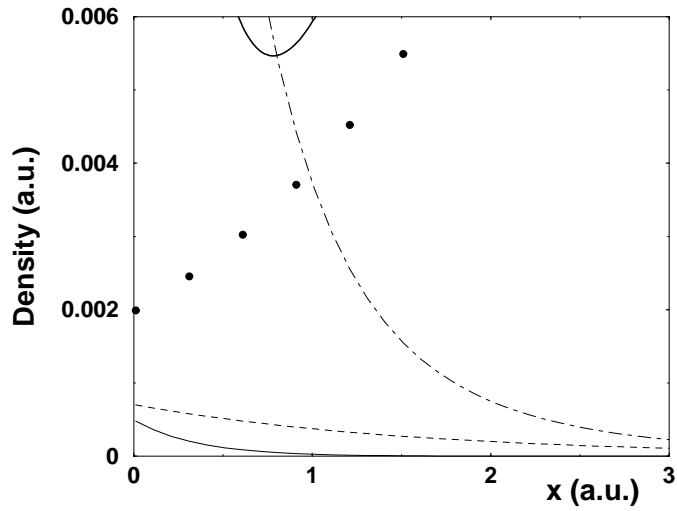}}
    \caption{Close-up of  Fig.~\ref{fig4} in the lower left corner.}
    \label{fig5}
\end{figure}

In Fig.~\ref{fig3} we show how the addition of an 
attractive delta potential, that creates a bound state, modifies 
the analogous contributions, since the bound state component has to be 
detracted from them. $P_{T,bound}$ is constant in time, since there 
is only one bound state, and it is no longer true that the total final 
probability equals $P_{T,+}(\infty)$; actually, we have 
$P_{T}(\infty)=P_{T,bound} + P_{T,+}(\infty)$. In the picture we have 
only shown the interference term between positive and negative 
momenta, which is the most relevant one in this case.
%
%
In order to show the applicability of 
expression (\ref{comp1}), we portray in Fig.~\ref{fig4} and 
Fig.~\ref{fig5} the squared moduli of the total amplitude and of different 
components of the amplitude  for a step potential with an attractive delta 
potential; that is, the potential is 
$V(x)=-U_{0}\delta(x)+V_{0}\theta(x)$. In this situation, additional 
to the positive and negative 
momenta components, we have evanescent waves and the 
contribution of the bound state. At the time of both pictures, ($t=5$ 
in atomic units), the evanescent waves are the major contributor to 
the wavefunction in the vicinity of the origin. It should be noticed 
though that interference terms are relevant in that region and 
elsewhere.

The magnitude of the component of negative 
momenta, as well as that corresponding to the bound state, is rather 
small relative to the positive momenta component of the amplitude 
modulus
squared, but it is nonetheless very important even for large distances 
because of the interference between components, as is clearly reflected 
in Fig.~\ref{fig4}. Fig.~\ref{fig5} is added to show more clearly the relative 
importance of the negative momenta and bound state components.

%
%
%

%

\acknowledgments{We acknowledge support
by Ministerio de Educaci\'on
y Cultura (Grants 
PB97-1482 and AEN99-0315), by The
University of the Basque Country (grant UPV 063.310-EB187/98), and 
the Basque Government (grant PI-1999-28).
A. D. Baute acknowledges an FPI fellowship by Ministerio de 
Educaci\'on y Cultura.}

\end{document}